\documentstyle[12pt,eucal]{article}

\def\be{\begin{equation}}
\def\ee{\end{equation}}
\def\l{\label}
\def\tm{$^{\mbox{\sc tm}}$}
\def\kcal{\mbox{ kcal}}

\def\O{\mbox{\rm O}}
\def\H{\mbox{\rm H}}
\def\C{\mbox{\rm C}}

\def\F{\mbox{\rm F}}

\begin{document}
\begin{titlepage}
\title{\Large\bf STRUCTURE AND COMBUSTION OF MAGNEGASES\tm}
\author{{\normalsize\bf R. M. Santilli$^1$ and A. K.
Aringazin$^{1,2}$}\\[0.5cm]
{\normalsize $^1$Institute for Basic Research,}\\
{\normalsize P.O. Box 1577, Palm Harbor, FL 64382, USA}\\
{\normalsize ibr@gte.net}\\
{\normalsize $^2$Department of Physics, Eurasian National University,}\\
{\normalsize Astana 473021 Kazakstan}\\
}

\date{\small June 3, 2000; Revised August 15, 2001\\ Final version December 9, 2001}

\maketitle

\abstract{In this paper, we study the structure and combustion of
magne\-gases$^{TM}$ (Patented and International Patents Pending),
new clean fuels developed by one of us (R.M.S.) [1], which are
produced as byproducts of recycling nonradioactive liquid
feedstock such as antifreeze waste, engine oil waste, town sewage,
crude oil, etc., and generally vary with the liquid used for their
production. A new technology, called PlasmaArcFlow\tm, flows the
waste through a submerged electric arc between conventional
electrodes. The arc decomposes the liquid molecules into their
atomic constituents, and forms a plasma in the immediate vicinity
of the electrodes at about 10,000$^o$ F. The technology then moves
the plasma away from the electrodes, and controls its
recombination into environmentally acceptable fuels. In fact. the
exhaust of magnegases show: absence of carcinogenic or other toxic
substances; breathable oxygen up 14\%; and carbon dioxide down to
0.01\%. The new fuels possess a new chemical structure first
identified by one of us (R.M.S.), which is characterized by
clusters of ordinary molecules and atoms under a new bond of
electromagnetic nature. These clusters constitute a new chemical
species different than the conventional molecules, since they are
stable at ordinary conditions while exhibiting no infrared
signature (other than those of conventional molecular
constituents), thus confirming that the bond is not of valence
type. In particular, the bonding due to the magnetic polarization
of the orbitals, from space to toroidal distributions, have
resulted to be dominant over electric effects. For this reason the
new chemical species is called ''Santilli's electromagnecules'' or
''magnecules''.

In this paper we study the novel magnecular structure of
magne\-gases$^{TM}$, and confirm that, when the original waste is
of fossil or organic type, magnecules are essentially constituted
by conventional molecules H$_2$, CO, CO$_2$,$H_2O$, plus
individual atoms of H, O, and C, as well as radicals such as HO, CH and C-O in
single valence bond, all these constituents being bonded together by strong
magnetic fields originating from the toroidal polarization of the orbits of
valence and other electrons. We then present, apparently for the first time, an
estimate of the binding energy of magnecules. We also study the combustion of
magnegases produced from liquid feedstock of fossil origin , and show that it
is fundamentally different than the combustion of any other fuel. Conventional
fuels are constituted by conventional molecules, and generally burn
according to only one dominant thermochemical reaction. In this
sense, the combustion of conventional fuels can be conceived as
the firing of a ''single stage rocket''. Magnegases\tm are instead
constituted by conventional molecules bonded into the new
magneclusters, thus having a multi-stage structure, and they
generally have a sequence of dominant thermochemical reactions. In
this sense, the combustion of magnegases can be referred to the
firing of a ''multi-stage rocket'', with different fuels in
different stages. In this paper we study, apparently for the first
time, the primary chemical reactions in the combustion of
magnegases of fossil origin. Our main conclusion is that fuels
synthesized under intense electric and magnetic fields can indeed
release energy in amounts much bigger than those predicted by
conventional chemical reactions. Since, in addition, the new fuels
can be produced everywhere, and have environmentally acceptable
exhausts, magnegases offer serious
possibilities to satisfy our ever increasing energy needs, as well
as to contain the alarming environmental problems caused by fossil
fuels. }
\end{titlepage}

\section{Introduction}

According to official data by recent U.S. Department of Energy
(DOE) release, about 74 billions barrels of fossil oil,
 corresponding to about four trillions gallons of gasoline, are
consumed in our planet per day in an estimated number of half a
billion cars, one million trucks, one hundred thousand planes,
plus industrial, agricultural, and military uses. The combustion
of such an enormous amount of fossil fuel per day has caused
increasingly alarming environmental problems, such as:

1) Measurable emission in our atmosphere of the largest amount of
carcinogenic and other toxic substances;

2) Measurable oxygen depletion in our planet due to the fact that
the sum of the oxygen emitted in the combustion exhaust and that
recycled by plants from the emitted CO$_2$ is smaller than the
oxygen used in the combustion (negative oxygen balance);

3) Measurable increase of carbon dioxide of dramatic proportions, with
potentially catastrophic climactic changes;

\noindent and other environmental problems, such as the production of
poisonous NOx.

As a result of studies initiated in the early 1980's at Harvard
University under DOE support (see review [1]), one of us (R.M.S.)
has recently developed a new technology, called PlasmaArcFlow\tm [1c] (Patented
and International Patents Pending), for the production of environmentally
acceptable combustible gases, called magnegases\tm (see the industrial web site
[2a] and the scientific web site [2b]).

The new technology is primarily conceived to recycle
nonradioactive liquid feedstock, such as antifreeze waste, engine oil
waste, town sewage, as well as crude oil, etc., by flowing the
liquid through a submerged electric arc between submerged electrodes within
a metal vessel. The arc decomposes the liquid into its atomic
constituents, ionizes the atoms, and forms a plasma in the
immediate vicinity of the tip of the electrodes at about
10,000$^o$~F. The new technology then moves the plasma away from
the arc, and controls its recombination into magnegases which
bubble to the liquid surface where they are collected with various
means, are pumped into conventional, low or high pressure, gas
storage tanks, and then used for metal cutting, heating, cooking,
automotive, and any other other fuel application [2].

It is evident that magnegases vary with the liquid used for their
production. However, magnegases produced from different liquid of
fossil fuel origin have a similar composition, thus admitting a
unified treatment. They also constitute magnegases with the
highest energy content, thus constituting the most interesting
class.

In fact, the plasma originating from the recycling of fossil waste
is essentially composed of mostly ionized H, O, and C atoms (plus
traces of other elements which generally precipitate as solids
and, as such, are ignored hereon). Due to the known affinity of C
and O, abundance of carbon in the plasmas assures the removal of
oxygen via the formation of CO, thus preventing an explosive
mixture of H and O. The PlasmaArcFlow then prevents the subsequent
oxydation of CO, thus eliminating unwanted percentages of CO$_2$.
The resulting gases are stable at ordinary temperature, do not
self-ignite even at high temperature, and need a flame or a spark
to ignite.

Therefore, magnegases of fossil waste origin are essentially
constituted by H$_2$ and CO, with traces of O$_2$, H$_2$O, CO$_2$.
The full combustion of such a mixture is manifestly without any
possible carcinogenic or other toxic substance, thus resolving the
environmental problem 1) of fossil fuels recalled above.

Since magnegases are fabricated, they can be produced in such a
way that there still remains a significant percentage of oxygen in
the exhaust. In this way, magnegases have a positive oxygen
balance, and resolve the second environmental problem of fossil
fuels.

Under a proper combustion studied in detail in this paper, carbon
is mostly released in its solid form, resulting in small
percentages of CO$_2$, thus resolving the third environmental
problem of fossil fuel.

The exhaust of magnegases produced from fossil waste is
constituted by: 50\% to 60\% H$_2$O (as water vapor); 10\% O$_2$;
3\% to 7\% of CO$_2$; the remaining components being inert
atmospheric gases.

Other features of the new technology important
for this study are the following. Since magnegases are internally
rich in oxygen, their combustion requires 30\% to 40\% of the air
intake needed for fuels of fossil origin. This implies a large
reduction of nitrogen participating in the combustion, with
consequential reduction of toxic NOx; for details on combustion
exhaust one can inspect [2c].

All types of  magnegases can be used in all existing internal
combustion (IC) engines, via the sole change of external
components, essentially dealing with the carburation, coil
voltage, and timing [2].

The equipment for the production of magnegases, known under the
name of {\it Santilli's hadronic reactors} (Patented and Patents Pending), are
commercially over-unity. In fact, independently certified measurements one can
inspect at the USMG laboratory in Florida, have established that, for the
reactor operating at atmospheric pressure, one unit of electric energy
calibrated at the panel produces at least three units of energy in
the gas, plus at least one unit of energy as heat in the liquid to
be recycled, resulting in the over-unity of 4 (bigger over-unities
are possible for the reactor operating at bigger pressures, due to
the decrease of the gas bubbles, although they will be ignored in
this paper).

As a consequence, hadronic reactors producing  magnegases are
essentially capable of tapping energy from liquid feedstock and from the carbon
electrodes, in essentially the same way as Fermi's nuclear reactors tap energy
from nuclei, with the difference that the former have been certified [2] to
have no harmful radiations and leave no harmful waste, while the latter
release harmful radiations and wastes.

Moreover, nuclear reactors are notoriously large and require large
protections due to the harmful radiations, thus being solely
usable in large plants far away from inhabited areas. By
comparison, since they need no shield for radiations, hadronic
reactors are small, typically having the dimensions of a desk
[2a]. As a result, magnegases can be produced everywhere needed,
thus avoiding the costs of transportation over large distance
which are typical for fossil fuels.

Since  magnegases are produced as byproducts of the recycling of
liquid waste or crude oil, since their production equipment is
commercially over-unity, and since they do not require
transportation of fuels over large distances, the cost of
magnegases is competitive over that of fossil fuels, of course,
when produced in sufficiently large volume. In fact, the two
incomes from the recycling of liquid waste and from the
utilization of the heat, plus the high efficiency of the reactors imply very
low operating costs.

The resolution of the alarming environmental problems caused by
fossil fuels, their possible use in any existing engine, burner,
or furnace, and their competitive cost over existing fuels, render
magnegases serious alternative sources of clean fuels, thus
deserving a serious study by the scientific community.

The reader should be also aware that, contrary to popular beliefs,
large uses of hydrogen are potentially catastrophic for our
planet, unless the hydrogen is produced via ecologically acceptable ways.
Gasoline produces trillions of cubic foot of CO$_2$ per day which is at least
in part recycled by plants into O$_2$. By comparison, hydrogen has one tenth
the energy content of gasoline, thus requiring at least ten times oxygen for
the same performance (say, the same m.p.h.). Moreover, all atmospheric oxygen
in the combustion of pure hydrogen is turned into water, thus being
permanently removed from our planet due to the notoriously high
cost of water separation.

A similar fate also holds for fuel cells, since they also operate by
burning hydrogen or similar environmentally unacceptable fuel. Besides,
fuel cells cannot be used in existing IC engines. Therefore, their uses
leaves completely unchanged the current deterioration of our
environment (since fuel cells are expected to be used only for {\it
new} cars, thus leaving unchanged existing cars).

On similar grounds, electric cars can admittedly improve local
urban environments, although they globally {\it increase}, rather
than decrease, environmental problems, since electric power plants
are known to be much more polluting than automobiles, whenever
they are of fossil or nuclear type.

In conclusion, the new combustible fuels magnegases constitute a
serious and real source of new clean fuels that are environmentally preferable,
not only with respect to fuels of fossil origin, but also with respect to
alternatives currently under study, such as hydrogen, fuel cells, and
electric cars.

It should be mentioned that entropy increases in the transition from a
solid state to a liquid or gas state. Therefore,the  high efficiency
of hadronic  reactors does not violate conventional thermodynamics laws.

The study of this paper is hereon restricted, specifically, to the
single magnegas produced from the recycling of antifreeze waste,
since this is the gas subjected to the largest number of
experimental measurements until now [1c]. However, our results are
easily extendable to other types of magne\-gases. The reader
should be aware that other types of magnegases, such as those
produced from the recycling of oil waste or crude oil, have an
energy content bigger than the magnegas produced from antifreeze
waste.

\section{The novel chemical structure of magnegases}

All types of magnegases (again referred to combustible gases
produced under an electric arc) possess a new chemical structure
first identified by one of us [1b], which is characterized by
clusters of ordinary molecules, radicals and atoms bonded together
by attractive forces of primary magnetic origin (see Ref. [1c], Chapter 8, for a
comprehensive presentation).

As established by numerous Gas Chromatographic, Mass Spectrometric
(GC-MS) tests under joint InfraRed Detectors (IRD),  these
clusters constitute a new chemical species different than that
characterized by conventional molecules (i.e., clusters with a
valence bond), since the magnegas clusters are stable at ordinary
conditions which remain unidentified among known molecules while
exhibiting no infrared signatures (other than those of
conventional molecular constituents), thus confirming that the
bond is not of valence type.

As well known in physics (although less emphasized in contemporary
chemistry), whether conventional or novel, stable clusters
detected by GC-MS equipment {\it cannot} exist without an {\it
attractive} force. The specific and concrete identification of the
attractive bond of electromagneclusters constitutes a central
aspect in the study of magnegas, with implications similar to
those of the valence for conventional molecules.

Extensive studies of this aspect have established that the primary
attractive force responsible for the electromagnecules is expected
to be due to the magnetic polarization of the orbitals of valence
and other electrons, from space to toroidal distributions.
Conventional quantum electrodynamics establishes the existence of
such a polarization whenever atoms and molecules are exposed to
intense magnetic fields, as it is the case in the vicinity of the
electric arc. Such a magnetic polarization creates magnetic
North-South polarities along the symmetry axis of the toroid,
which permit the stacking of atoms and molecules one after the
other [1c].

The above analysis has permitted a quantitative interpretation of
the detection via GC-MS/IRD of clusters possessing molecular
weight all the way to 1,000 a.m.u. in a {\it light }gas, such as
magnegas from antifreeze waste, whose highest molecular weight
should be 44 a.m.u for CO$_2$.

Electric contributions are expected to exist but their study has
not been conducted to date. Conventional molecules are mostly
preserved in the new clusters (as established by the preservation
of their IR signature). Therefore, valence electrons cannot
contribute to the bonding force of the new clusters. Ionic
contributions, even though unquestionably present, are notoriously
unacceptable for a bond, because they carry the same charge, thus
resulting in a {\it repulsive}, rather than attractive force.
Electric polarization (essentially due to a deformation of the
charge distribution of atoms) are also unquestionably present,
although they are notoriously unstable and weak, while their
geometry does not lend to large clustering when acting alone.

As a result of the expected dominance of magnetic over electric
contributions, the new chemical species composing magnegas is
called ''magnecules'' or ''magneclusters'' [1b]. The same
terminology will be adopted in this paper hereon.

When originating from fossil or organic waste, numerous GC-MS/IRD
scans [1c] have established that gas magnecules are essentially
constituted by conventional molecules H$_2$, CO, CO$_2$, H$_2$O,
plus individual atoms of H, O, and C, as well as radicals such as
HO.

It is evident that the bonding of atoms and molecules into new
clusters constitutes new means for storing energy in a combustible
gas, in addition to the conventional valence means in ordinary
fuels. Therefore, magnecules have a primary relevance for the
study of the combustion of magnegas. It is also evident that a
deeper understanding of the magnetic origin of the new clusters in
magnegas will permit an increase of their energy content, since
the size and strength of magnecules is directly proportional to
the magnetic field used for its formation.

Magnegases have additional new means for storing energy deep into
the structure of conventional molecules. Recall that any peak in
the IR signature of conventional molecules denotes a corresponding
valence bond. For instance, an IR signature with two peaks for CO
denotes the existence of a valence bond of the type C=O.

Numerous independent measurements on magnegas [1c, 2a] have
established that the IR of conventional molecules is ''mutated''
with the presence of new additional peaks. Since all valence
electrons are generally used in the molecules, the new IR peaks
can only be explained via the creation of new internal bonds of
non-valence type. Again, the magnetic origin of these new bonds
has resulted to be plausible, and essentially consists in the
toroidal polarization of the orbitals of internal electrons under
external, strong magnetic fields. For the case of C=O, these
internal toroidal polarizations imply the creation of the pair of
magnetic polarities $(North-South)\times (North-South)$. The
creation of internal magnetic bonds then alters the entire
thermochemical horizon, beginning with {\it new} values of binding
energies for {\it conventional} molecules.

It is evident that a deeper understanding of the new internal bonds in
conventional molecules may permit a corresponding increase of energy
storage in fuels, since, again, their number and strength depends on
the intensity of the external magnetic field used for their creation.

A technical understanding of this paper also requires a knowledge
of various additional anomalous features of magnegas, such as
their {\it anomalous adhesion} (which occludes feeding lines with
small sectional areas usually acceptable for conventional gases,
thus preventing the passage of large magnecules to be detected);
{\it mutations} (generally referred to macroscopic alterations in
time of magnecules under the same conditions following break-down
due to collisions and subsequent recombinations); {\it alteration
of physical characteristics} (such as macroscopic changes in
average molecular sizes, density, viscosity, etc.); and other
anomalies.

In this paper we study the above anomalous new means for storing
energy in magnegas and introduce, apparently for the first time,
an estimate of the binding energy of magnecules.

\section{The novel combustion of magnegases}

Predictably, the new chemical structure of magnegases implies that
its combustion is fundamentally different than the combustion of
conventional fuels, possessing a conventional chemical structure.
A conventional fuel is constituted by a given conventional
molecule, and generally burns according to thermochemical
reactions. In this sense, the combustion of a conventional fuel
can be referred to the firing of a ''single stage rocket''.

Magnegases are instead constituted by conventional molecules
bonded into the new magneclusters, thus having a multi-stage
chemical structure. In addition, as we shall see, magnegases
generally have a sequence of dominant thermochemical reactions. In
this sense, the combustion of magnegases is here referred to the
firing of a ''multi-stage rocket'', with different fuels in
different stages.

In this paper we study, apparently for the first time, the primary
chemical reactions in the combustion of magnegases of fossil
origin. Our main conclusion is that fuels synthesized under
intense electric and magnetic fields can indeed release energy in
amounts much bigger than those predicted by conventional chemical
reactions.

To initiate this study, consider the simplest possible magnegas,
that produced from ordinary tap water and a carbon electrode via
an electric arc having 1,000 A, 40 V DC, and the PlasmaArcFlow set
to produce a gas with about  45\% H$_2$ and 45\% CO, the remaining
percentages being composed of H$_2$O, CO$_2$, and O$_2$. The
primary effect of the arc is the dissociation,
        H$_2$O $\to$ 2H + O.
Owing to the excess of carbon in the plasma we then have the known
reaction,
        2H + O + C $\to$ H$_2$ + CO,
which explains in part the reason why H$_2$ and CO are observed in
essentially equal percentages.

Consider then the elementary sequence of reactions
\be\l{channel}
        \H_2\O + \C \to 2\H+\O+\C \to \H_2 + \C\O.
\ee
Let us calculate the energy balance for the reaction channel.
Irrespective of possible intermediate reactions, the minimal
energy needed will be the same, and it can be calculated via to
the energy balance:
\be\l{balance}
        E[producing] = E[\H_2] + E[\C\O] - E[\H_2\O] - E[\C],
\ee
where $E[\H_2]$, $E[\C\O]$, $E[\H_2\O]$, and $E[\C]$ denote ground
state energies.

Since the main channel is (\ref{channel}) one can suppose that,
microscopically, the production of H$_2$ and CO occurs as follows.
Immediately after a given H$_2$O molecule has been bombarded by
the electrons of the arc, it dissociates. Then, near to the arc
(in a lower temperature region, to which the atoms are moved by
flow), when the C atom occurs near the H+O+H complex, the O atom
captures the C atom while the two H atoms composes the H$_2$
molecule.

Processes of ionization of H, O, C, H$_2$O, OH, and CO, under the
effect of the arc, as well as excited states and radiation, are
not considered here, for simplicity.

The process of creation of H$_2$ and CO are thus running in
parallel, with several similar processes occurring in neighborhood
of a given H$_2$O molecule. So, these elementary processes occur
at the same conditions (temperature, pressure, external
electromagnetic field produced by the arc and neighbor electrons,
ions and atoms).

This is the point where and when the production of magnecules in a
gaseous phase takes place in part. The magnecules may, therefore,
consist of several H$_2$ and/or CO molecules, plus individual H, O
and C atoms and OH radicals, as well as C-C and C-H dimers, bonded
to each other by chains of North-South polarities created by the
toroidal polarization of the orbits of valence and other electrons
[1b].

As the PlasmaArcFlow moves the magnecules away from the high
temperature region, magnegas cools down, and bubbles to the
surface of the liquid, where it is collected, for storage and use.

Due to the detected rather big mass of magnecules (up to 1000
a.m.u.) at ordinary temperature, it seems that CO (with mass 28
a.m.u) is the heaviest molecular constituents in large percentage,
followed by the heavier CO$_2$ molecule (with mass 44 a.m.u.),
although in much smaller percentage.

Indeed, it is hard to imagine that, the peak with 416 a.m.u
detected in  magnegas by GC-MS/IRD tests can be due to a magnecule
consisting of 208 H$_2$ molecules. Such a cluster can probably not
survive at room temperature due to its huge size, as well as
linear character of its bond. In fact, a magnecule of 208 bond
lengths of H$_2$ has the linear length $208\times 1.4 \approx 300$
bohrs, thus being excessively long for stability.

The same weight of 416 a.m.u. can be better represented by only 15
CO molecules, in which case the linear size is $15\times 2 = 30$
bohrs. In addition, CO has plenty of electrons (some of which are
unpaired), in contrast to H$_2$. These electrons might be
responsible for magnetic polarizations as needed in the magnecule,
in accord to the observed infrared spectrum of CO [1c].

However, the magnecules consisting solely of CO molecules can not
probably explain why magnegas possesses an anomalously high energy
release since H$_2$ is a combustible component too, which can make
a great contribution to the total energy released. Thus, most
likely, the magnecules of 416 a.m.u. consists of both CO and
H$_2$, e.g., with 14 CO and 12 H$_2$ molecules. The latter
assumption explains the magnecule mutation of two a.m.u.

In summary, the elementary reaction channel (\ref{channel}) for
producing H$_2$ and CO, can be extended to
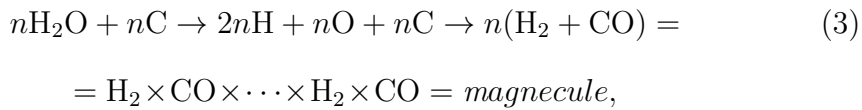
\be\l{channelmag}
n\H_2\O + n\C \to 2n\H + n\O + n\C \to n(\H_2 + \C\O) =
\ee
$$
= \H_2\!\times\!\C\O\!\times\!\cdots \!\times\! \H_2 \!\times\!
\C\O = {\textit magnecule},
$$
where the final system is a type of magnecule consisting of equal
number $n$ of bonded H$_2$ and CO molecules (not $n$ separate
molecules H$_2$ and $n$ separate molecules CO), where "$\times$"
denotes the magnetic bond.

The further assumption of 14 CO, 11 H$_2$ and 2 isolated H atoms
then explains the magnecule mutation of only one a.m.u. This
latter assumption is not considered in these introductory remarks
for simplicity.

The following remark is in order. We should note that the magnetic
origin of the bond is not critical here, so "$\times$" can be
thought of as some kind of bond which is weaker than the bonds H-H
and C=O. So, the model, which we use here, is in essence rather
general, and can be used for other possible interpretation of the
attractive force permitting the existence of the anomalous
clusters.

Under the above assumptions, the energy needed for production of
magnecules is
\be\l{balancemag}
        E[production] = E[magnecule] - nE[\H_2\O]-nE[\C],
\ee
where $E[magnecule]$, $E[\H_2\O]$, and $E[\C]$ are ground state
energies, and we assume, again for simplicity, the magnecule as
consisting of equal number $n$ of CO and H$_2$ molecules.

The above amount of energy may be indeed different from
(\ref{balance}) because the ground state energy of the magnecule,
$E[magnecule]$, is evidently not equal to $nE[\H_2] + nE[\C\O]$
due to the binding energy of magnecule, eventual deformation of
orbitals, and other internal electronic effects. This is the first
case where the anomalous efficiency of the PlasmaArcFlow reactor
may originate.

\section{Combustion of H$_2$ and CO in equal percentage}\l{H2CO}

In this section we present elementary calculations of reaction
heats, for the combustion of H$_2$, CO in a 50\%+50\% mixture. The
binding energies of interest are presented in Table~1 [4], where
the complete dissociation of gas states of reagents to gas states
of product atoms is assumed at temperature $T=25$ C.

\begin{table}
\begin{center}
\begin{tabular}{|cc||cc|}
\hline
Diatomic molecules &&Diatomic molecules &\\
\hline
H$-$H              &104.2 & C=O & 255.8\\
\hline
O=O              &119.1 & N$\equiv$N &192.0\\
\hline
\hline
Manyatomic molecules &&Manyatomic molecules &\\
\hline
C$-$O                 &85.5    & O$-$H & 110.6\\
\hline
C=O in CO$_2$       &192.0   & O$-$O & 35\\
\hline
C=O in formaldegide &166     & C$-$H & 98.7\\
\hline
C=O in aldegides    &176     & C$-$C & 82.6\\
\hline
C=O in ketons       &179     & C=C & 145.8 \\
\hline
C=N                 &147     & C$\equiv$C &199.6\\
\hline
\end{tabular}
\caption{Binding energies, kcal/mole. $T=25$C.}
\l{Table1}
\end{center}
\end{table}

{\bf Combustion of H$_2$:}     H$_2$ + O$_2$/2 $\to$ H$_2$O. \\
The reactions are:
\be \H_2 \to \H + \H - 104.2,\ee
\be \O_2/2 \to  (\O + \O)/2 - 119.1/2,\ee
\be \O + 2\H \to \H_2\O + 2\times110.6.\ee
So, the energy balance is
$2\times110.6 - 119.1/2 - 104.2 = 57.5$.
Thus,
\be \H_2 + \O_2/2 \to \H_2\O + 57.5 \kcal.\ee

{\bf Combustion of CO:}     CO + O$_2$/2 $\to$ CO$_2$. \\

The reactions are:
\be \C\O \to \C + \O - 255.8,\ee
\be \O_2/2 \to  (\O + \O)/2 - 119.1/2,\ee
\be \C + 2\O \to \C\O_2 + 2\times192.\ee
So, the balance is
$2\times192 - 119.1/2 - 255.8 = 68.7.$
Thus,
\be\l{COOCO}
\C\O + \O_2/2 \to \C\O_2 + 68.7 \kcal.
\ee

{\bf Combustion of the 50\%+50\% mixture of H$_2$ and CO.}\\
The balance is $2\times110.6 + 2\times192 - 119.1 - 255.8 - 104.2
= 126.1$ kcal per 2 moles (1 mole of H$_2$ and 1 mole of CO). For
1 mole of the mixture, we then have $126.1/2 = 63$ kcal. Thus,
\be
(\H_2 + \C\O) + \O_2 \to \H_2\O + \C\O_2 + 63 \kcal.
\ee
So, the mixture of H$_2$ and CO gives bigger reaction heat than
100\% H$_2$ gas.

Note that, since 57.5 kcal/mole is equal to 300 BTU/cf, the
50\%+50\% mixture of H$_2$ and CO gives about 63 kcal/mole = 330
BTU/cf.

\section{Combustion of acetylen}\l{Acetylen}

In this section we study the thermochemical properties of
acetylene because useful for the study of magnegas when used for
metal cutting [2a].

Pure acetylene (which {\it is not} the acetylene used in the metal
cutting industry) is a liquid with the linear molecule C$_2$H$_2$
= H-C$\equiv$C-H,   boiling temperature of $T= - 84$C, and melting
temperature of $T= - 81$C. The heat released in the combustion of
acetylene is  311~kcal/mole, at temperature $T=25$C.

The combustion of acetylene in pure oxygen occurs via the reaction
\be\l{acet}
\C_2\H_2 + (5/2)\O_2=2\C\O_2 (\mbox{gas})+\H_2\O (\mbox{liquid})
+311\mbox{ kcal/mole},
\ee
holding at the flame temperature of about $T=2800$C.

It is interesting to note that, e.g., ethane, $C_2H_6$, has a
bigger reaction heat of 373 kcal/mole, but it holds at the smaller
flame temperature $T \ll 2800$~C due to the reaction,
\be
\C_2\H_6+(7/2)\O_2 = 2\C\O_2 (\mbox{gas})+3\H_2\O (\mbox{liquid})
+373\mbox{ kcal/mole},
\ee
producing three times water of the preceding reaction. A bigger part of
the total released energy is spent in water evaporation, by decreasing
the flame temperature. It follows that ethane is a combustible gas less
efficient than acetylene for metal cutting.

Magnegas is currently in regular production and sales for metal
cutting [2a]. Its daily uses in this field has established that
magnegas is more effective than acetylene in metal cutting,
because: 1) the pre-heating time is about half; 2) the cutting
speed is almost double; and 3) the combustion exhausts are
dramatically cleaner than those of commercially sold acetylene; 4)
there is no ''back-flash'' (reflection of flame in rusty
surfaces); and 5) the cut is manifestly cleaner. Magnegas is also
much safer because stable, as compared to the notoriously unstable
commercial acetylene; it can be produced anywhere needed with a
desk-size equipment, thus avoiding transportation altogether; and
it is cheaper than acetylene.

The better efficiency of magnegas compared to acetylene is
primarily due to the combination of a bigger flame temperature and
energy density, which reduce the pre-heat time and increase the
cutting speed.

In fact, the flame temperature for pure H$_2$ is about $T=3100$ C,
so that 50\% of H$_2$ and 50\% of CO in magnegas could provide a
flame temperature bigger than that of acetylene.

An additional reason for the bigger efficiency of magnegas in
metal cutting is due to the fact that only a portion of oxygen is
of atmospheric origin, while the remaining portion originates
from: 1) free O-atoms in magnecules; 2) the presence in magnecules
of HO-dimers; and 3) a possible dissociation of C$\equiv$O, C=O,
and C--O groups, which we expect to be present in magnegas. These
features are confirmed by the dramatic decrease of air or oxygen
for the burning of magnegas as compared to acetylene, as well as
by observed carbon residues.

It should be noted that the value of the reaction heat of 311
kcal/mole, has been measured for the water in liquid state (see
Eq. (\ref{acet})), while the combustion heat of magnegas has been
studied until now only for water produced at the vapor state.
Therefore, the comparison of combustion heat of magnegas with that
of acetylene should require both combustion heat measured under
the same conditions, i.e., both combustions being measured for
water produced either at the liquid or at the gas state, since the
evaporation heat of water is considerable (10.4~kcal/mole at
$T=25$~C). This occurrence is an additional reason for the
difficulties encountered until now in achieving a scientific
measurement of the energy content of magnegas [2a].

Additional difficulties in measuring the energy content of
magnegas are due to difficulties in achieving its complete
combustion in air. This is due to the anomalous means of storing
energy, some of which is stored deep into the structure of
conventional molecules, which evidently requires special
conditions for complete combustion.

The Konovalov's equation permits an estimate of the combustion heat,
\be\l{Kon}
\Delta H = -(47.02m+10.5n+x) \mbox{ kcal/mol},
\ee
where $m$ is number of O atoms used for complete combustion, $n$
is number of moles of the water produced, and $x$ is the
correction coefficient, which is a characteristic constant for
each gas. Therefore, final measurements on the magnegas combustion
will permit the identification of its characteristic constant $x$.

\section{Combustion of magnegas}\l{magnegas}

It is evident that the combustion of magnegas requires the
oxidation, first, of magnecules, and then that of conventional
molecules H$_2$ and CO. Therefore, the preceding energy
calculations for a gas with 50\% H$_2$ and 50\% CO do not apply
for a magnegas with the same conventional chemical composition.
Additional novelties occur for other types of magnegases, such as
that produced from the recycling of antifreeze waste, whose energy
content has been conservatively estimated to be of the order of
900 BTU/cf [2a].then, the bonding structure of magnecules plays a
key role in understanding the energy content of magnegas.

To begin, the known reactions
        H$_2$ + O$_2$/2 $\to$ H$_2$O and
        CO + O$_2$/2 $\to$ CO$_2$,
should be replaced by the reaction
\be
  magnecule + n\O_2 \to m\H_2\O + k\O_2+ l\C\O_2 + ... + \Delta\mbox{
kcal},
\ee
which may give increased energy released per each H$_2$ molecule.
Here, $n$, $m$, $k$, $l$, ... are numbers, and the original
magnecule is assumed to consist of both H$_2$ and CO molecules.

The energy balance for the combustion of magnecule is then given by:
\be
        E[combustion] = mE[\H_2\O]+ kE[\O_2]+ lE[\C\O_2] + ... -
E[magnecule],
\ee
where $E[H_2O]$, $E[O_2]$, $E[CO_2]$, ... are ground state energies of
the molecular constituents, and $E[magnecule]$ is ground state energy
of the the original magnecule.

A way to calculate this energy balance is to use dissociation
energy of the magnecule, $D[magnecule]$. However, we should note
that $D[magnecule]$ is different for magnecules of different mass
and composition.

For chemical reactions, one should take into account the value of
the reaction constant $K$. For example, for the reaction
$\H_2+\O_2/2 \to \H_2\O$ ($\Delta H=-57.5$ kcal), the reaction
constant is very big, $K=10^{40}$ at $T=25$ C, thus indicating an
almost total combustion of H$_2$ gas at $T=25$ C.

The general rule is that, for all {\it highly exothermic}
reactions (typically, with $\Delta H<-15$ kcal/mol), the reaction
constant is of high value. The opposite direction of the reaction,
$\H_2+\O_2/2
\leftarrow \H_2\O$, is realized only at very high temperatures, at
which
$K<1$. The value $K=1$ means equilibrium of a reaction, while $K<1$
means that a reaction runs in opposite direction. In general, the
relation between the reaction heat, $\Delta H$, and the reaction
constant, $K$, is as follows:
\be
-2.303RT\, \mbox{log} K =\Delta G,
\ee
where
\be\l{G}
\Delta G \equiv \Delta H - T\Delta S,
\ee
$R=1.986$ cal$\cdot$K$^{-1}\cdot$mol$^{-1}$, $T$ is temperature in
Kelvins, and $\Delta S$ is the entropy of the reaction. The latter
is numerically big if the initial reagents have molecular
structures more ordered than the end products, i.e. there is an
{\it increase} of entropy $S$ during the reaction.

The above outline on the reaction constant and reaction entropy
helps us to conclude that the combustion of magnegas is
characterized by a very high value of the reaction constant
(perhaps even bigger than $K=10^{40}$ at $T=25$C). In fact, the
combustion of magnegas is a highly exothermic reaction, and the
magnecules have a structure much more ordered than the product of
the combustion. Therefore, during the combustion of magnegas we
have a large increase of the entropy $\Delta S>0$. These two
factors lead to very high value of the reaction constant $K$ for
the combustion of magnegas.

The table values of $\Delta H$ and $\Delta G$ are given at normal
conditions ($T=25$C, $p=$ 1 atm). However, $\Delta G$ is a function of
the temperature. For most elements, $\Delta G$ of oxidation reactions
(linearly) increases with the increase of the temperature. Thus, the
resulting oxides are less stable at high temperatures than at low
temperatures (typical example is $H_2O$ which dissociates at very high
temperatures).

It is interesting to note that the oxidation of carbon to carbon
monoxide, e.g., C + CO$_2$ $\to$ 2CO, is almost the only oxidation
reaction for which $\Delta G$ {\it decreases} with the increase of
the temperature. Here, the number of moles increases about twice
during the reaction. As a result, the entropy greatly increases,
$\Delta S>0$. Therefore, the CO molecule is {\it more} stable at
high temperatures than at low temperatures (for example, it is
about twice more stable at 3000C than at 1000C).

Another ecologically very important aspect in the combustion of
magnegas is therefore the {\sl reduction of} CO$_2$ {\sl via the
oxidation of carbon atoms present in magnecules}, {\sl and its
subsequent dissociation as in Eq. (9) to release the oxygen needed
for the burning of} H.

Since the stability of CO increases with the temperature, {\sl a
better quality of the exhaust is reached at lower original
temperatures of magnegas}. This result should be compared with the
opposite occurrence for natural gas and for other fuels, which are
generally {\it pre-heated} prior to combustion.

Another important characteristics of a reaction is the reaction
rate. Various tests [2a] have show that the combustion of
magnecules is faster than the combustion of their molecular
constituents. Santilli-Shillady isochemical models of molecular
structures permits the following understanding of this additional
anomaly.

In their natural conventional, and non-polarized states, H$_2$ and
O$_2$ molecules have the usual (spherical) shape due to rotations.
However, an inspection of the isochemical model of the water shows
that such configurations are not suited for the reaction of H and
O into H$_2$O. In particular, the orbitals of H$_2$ require a
toroidal configuration as a condition for their bonding to the
oxygen, a similar occurrence holding for the oxygen too.

It then follows that {\sl magnetically polarized molecules of
hydrogen and oxygen have a bigger reaction rate than the same
molecules in un-polarized conditions, since they have a
distribution of the valence electrons more suitable for the
reaction itself}. Evidently, a bigger reaction rate implies a
bigger power.

Moreover, the combustion of a magnecule consisting of H$_2$ and
CO, does not require the necessary previous dissociation of the
O$_2$ molecule (O$_2 \to$ 2O $-$ 119.1 kcal), because each O-atom
in a magnetically polarized O$_2$ molecule is ready for the
combustion. Therefore, the magnecular structure acts as a kind of
catalysis, in which both O-atoms of the O$_2$ molecule start to
react with the nearest pair H$_2\times$ H$_2$, or H$_2\times$ CO,
or CO$\times$CO almost simultaneously; see Fig.~1.

This occurrence also implies that less amounts of external energy
is needed to activate the reaction, resulting, again, in an
anomalous energy release in combustion. Usually, the activation
energy is supplied by heat.  Therefore, we can conclude by saying
that the combustion of magnegas can be initiated at smaller
temperature, in comparison to that of the simple mixture of H$_2$
and CO gases. The reduction of pre-heating time of about one-half
by magnegas as compared to that for acetylene, is in agreement
with the above interpretation.

\begin{figure}[ht]
\begin{center}
\unitlength=0.5mm
\begin{picture}(50,50)
\put(9,40){\line(0,0){7}}
\put(11,40){\line(0,0){7}}
\put(7,31){O}
\put(7,51){O}
\put(36,40){\line(0,0){1}}
\put(36,42){\line(0,0){1}}
\put(36,44){\line(0,0){1}}
\put(36,46){\line(0,0){1}}
\put(30,31){CO}
\put(30,51){CO}
\put(17,34){\line(3,0){2}}
\put(20,34){\line(3,0){2}}
\put(23,34){\vector(3,0){5}}
\put(17,54){\line(3,0){2}}
\put(20,54){\line(3,0){2}}
\put(23,54){\vector(3,0){5}}
\end{picture}
\end{center}
\vskip -20mm \caption{Collision of O$_2$ and a pair of correlated
CO molecules.}
\end{figure}
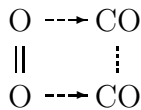

Due to the presence of magnecules viewed as heavy complexes of
H$_2$ and CO, one cubic foot of magnegas contains in fact {\sl
bigger number} of H$_2$ and CO molecules than it is for the
respective simple mixture (i.e., without the clusters) of these
two gases, at the same temperature and pressure. Clearly, magnegas
is far from being an ideal gas. For example, for ideal gases the
number of molecules is $N_A = 6.02 \cdot 10^{23}$ per 22.414
liters of volume, at normal conditions ($T=0$ C, $p=1$ atm). This
is not the case for real gases (e.g. CO$_2$), and especially for
the gases containing macroscopic percentage of particles of huge
mass, 400...800 a.m.u., (and therefore big effective size) which
are interacting with each other. Thus, the bigger number of H$_2$
and CO molecules per cubic foot of magnegas can be considered as
one of the simple reasons of high energy content of magnegas which
has been measured and calculated per cubic foot.

The above note implies that more experimental study of the
thermodynamics of magnegas, viewed as a real gas, is needed. For
example, measurement of the critical parameters ---pressure $p_k$,
volume $V_k$, and temperature $T_k$--- and the Jole-Tomson effect
is of primary interest here. Also, we need in measurement of the
energy released of magnegas combustion not per cubic foot but per
mole, to obtain a value of the reaction heat, which we then can
compare to the standard values of the reaction heats, given in
kcal/mol, at $T=25$ C and normal pressure.

Note that oxidation of C can result in various compounds, CO,
CO$_2$, C$_3$O$_2$, C$_5$O$_2$, and even C$_{12}$O$_9$. The
reaction constant $K$ varies with the reaction temperature so that
at $T=600...700$ K the product is almost only CO$_2$ while at
about $T=1300$ K the product is almost only CO. At low
temperatures, the resulting carbon monoxide does not oxidize
further to carbon dioxide CO$_2$ despite the fact that this is
thermodynamically much preferable since the activation energy of
the reaction $\C\O+\O_2/2 \to \C\O_2$ is high. Also, a lack of air
(oxygen) supports producing of CO.

Below, we consider the reaction,
\be\l{COwater}
\C\O + \H_2\O \to \C\O_2 + \H_2 + 9.7 \kcal,
\ee
(low exothermic, experimental value is $\Delta H = -9.7$ kcal/mol,
$K= 10^5$, at $T=25$ C, gaseous states, constant pressure), which
also can take place. The theoretical value of the reaction heat is
due to
 $\C\O \to \C + \O - 255.8$,
 $\H_2\O \to \O + 2\H - 2\times110.6$,
 $\C + 2\O \to \C\O_2 + 2\times192$,
 and
 $\H + \H \to \H_2 + 104.2$.
 So, the balance is $104.2 + 2\times192 -
2\times110.6 - 255.8 = 11.2$. Thus,
\be\l{COHO}
\C\O + \H_2\O \to \C\O_2 + \H_2 + 11.2 \mbox{ kcal}.
\ee
The obtained theoretical value, 11.2 kcal, is a bit bigger than
the above experimental value, 9.7 kcal,  but it is acceptable
within the provided accuracy.

The reaction constant of this reaction decreases with the increase
of temperature and is about $K=1$ at temperature $T=830$ C so that
at higher temperatures the reaction goes in opposite direction,
\be\l{COHOb}
\C\O + \H_2\O \leftarrow \C\O_2 + \H_2, \quad\quad T>830 \mbox{C}.
\ee
At lower temperatures, $T=700...830$ C, we should expect that
magnegas exothermically reacts with water due to (\ref{COwater}),
thus decreasing percentage of CO and increasing percentage of
CO$_2$ and H$_2$ in magnegas. At room temperatures, CO does not
react with water because of the lack of activation energy needed
to initiate the reaction.

Using the reactions
\be\l{COCO2}
\C+\O_2 \to \C\O_2 + 94.5 \mbox{ kcal/mol},
\ee
\be\l{COCO}
2\C+\O_2 \to 2\C\O + 53.3 \mbox{ kcal/mol},
\ee
we obtain the endothermic reaction,
\be\l{CCOCO}
\C+\C\O_2 \to 2\C\O - 40.8 \mbox{ kcal/mol},
\ee
which runs from left to right at high temperatures and/or high
pressures, and it runs from right to left at low temperatures.
This reaction is important in {\sl optimization of the
PlasmaArcFlow reactor to produce magnegas with low percentage of
CO$_2$, and also to cool down the liquid}. Also, competition of
the reactions (\ref{COCO}) and (\ref{CCOCO}) is important in {\sl
optimizing the magnegas combustion}. More studies are needed to
identify optimal range of temperatures and pressures, for the
reactions (\ref{COOCO}), (\ref{COwater}),
(\ref{COCO2})-(\ref{CCOCO}), to get desired content of magnegas,
content of the exhaust, and its energy characteristics.

It is worthwhile to note that the mixture of CO and H$_2$, used as
a gas fuel and in a synthesis of hydrocarbons, can be produced by
the reaction of water vapor flow with overheated carbon (coal).
The PlasmaArcFlow reactor can be considered as an alternative
method to produce such a mixture, which is used in very important
process of {\it the catalytic synthesis of various hydrocarbons}.

Also, it is known that the mixture of 25\% (volume) of CO, 70\% of
N$_2$ (nitrogen), and 4\% of CO$_2$ is used as a gas fuel, with
the energy released about 1000~kcal per cubic meter.

\section{Structure of magnecules}\l{Structure}

In this section, we discuss on possible structures of magnecules
and the origin of the bonds in magnecule.

Currently, not much is known on the physical-chemical
characteristics of magnegas besides the basic chemical content,
energy released, infrared spectroscopy and gas-chromatography
mass-spectroscopy data, and some laboratory and industrial tests
certifying its anomalous energy characteristics and combustion
properties \cite {1,2,3}. Particularly, the IR/GC-MS data of
magnegas at room temperatures, $T=10...20$C, show that the
(fragments of) magnecules have masses up to 1000 a.m.u.,
(typically, 400...800 a.m.u. for the biggest mass in macroscopic
percentage), and by the mass-spectra they are not recognized as
known molecules by computer search among more than 100,000 known
species. Also, the infrared spectra reveal sharp peaks identified
as CO and CO$_2$ spectra, and a big number of very small peaks,
for the IR/GC-MS scanned masses 40...1000 a.m.u. Here, the
presence of CO infrared peaks is of somewhat strange character
since $M[\C\O]=28$ a.m.u. is out of the scanned mass range, $M>40$
a.m.u., so that none of separate CO molecules have been analyzed
by the infrared spectrometer. This can be interpreted as that CO
molecules are inside the magnecules, or some other effects
simulating appearance of the CO peaks take place.

The presence of CO$_2$ infrared peak, $M[\C\O_2]=44$ a.m.u., is
not unexpected and can be interpreted as that there are some
percentage of CO$_2$ in magnegas (as it is indeed the case due to
chemical analysis), and/or CO$_2$ molecules are also inside the
magnecules. The small infrared peaks have not been recognized by
computer search as the peaks corresponding to known molecules.
They could be interpreted as the vibrational-rotational spectrum
of the magnecules. We refer the reader to Refs.\cite{1,2} for the
other interesting information on magnegas, and its applications.

Conduction of precise tests on the basic characteristics of
magnegas and the detected magnecules is much important.
Typically, they are:\\
(Averaged) Molecular mass;\\
Density;\\
Boiling temperature;\\
Melting temperature;\\
Critical temperature at which the detected magnecules of magnegas
are completely destroyed (temperature stability of the combustion
properties);\\
Oxidation states;\\
Electronegativity;\\
Electrical conductivity;\\
Thermal conductivity;\\
Heat of vaporization;\\
Specific heat capacity;\\
First ionization potential;\\
Crystal structure;\\
Acid/Base properties;\\
Molecular volume;\\
Nuclear magnetic resonance (NMR) spectrum;\\
Raman spectrum;\\
Magnetic susceptibility;\\
Dissociation energy of magnecules; \\
Electric charge of magnecules; \\
Magnetic moment of magnecules;\\
Roentgen-structural analysis of magnecules.\\

The NMR spectrum of magnegas is one of the best ways to see
experimentally that there is a kind of magnetic bonds in
magnecules if H atoms are present in magnecules.  NMR test will
give us information on the strength of magnetic field inside the
H-atoms, to an extremally high accuracy.  So, if H$_2$ molecule is
polarized in magnecule, even to a very small degree, then NMR
spectrum would show this clearly due to high sensitivity provided
by modern NMR facilities. Also, this would give some experimental
grounds, at a microscopic level, to develop a realistic model for
magnecules.

The NMR chemical shift, $\delta$, is an entity characterizing
intensity of magnetic field experienced by the nucleus of H-atom
(proton). It is measured in ppm, in respect to the shift of
thetramethilsilane (TMS), (CH$_3$)$_4$Si, which is taken as the
standard zero chemical shift, $\delta[\mbox{TMS}]=0$ ppm. For
example, $\delta[\mbox{bare proton}]=31$ ppm,
$\delta[\H\mbox{-atom}]=13$ ppm, $\delta[\H_2\mbox{ gas}]=4.4$
ppm, $\delta[\H_2\O\mbox{ at T=0 C}]=5.4$ ppm,
$\delta[\H_2\O\mbox{ vapor}]=0.7$ ppm. $\delta[\mbox{HBr}]=-4.2$
ppm. The above differences in values of $\delta$ are due to
different degrees of diamagnetic screening of the proton caused by
electrons; less value of $\delta$ means that bigger diamagnetic
screening takes place.

For the case if H$_2$ is not in magnecules of magnegas, we would
see the NMR peak at about $\delta[\H_2\mbox{ gas}]=4.4$ ppm. For
the case if H$_2$ is in magnecules of magnegas, we would see the
NMR peak at some other value of $\delta$. In addition, we would
see the {\it number} of H-atoms in magnecule, and, if these
H-atoms are characterized by different electronic (magnetic)
environment inside the magnecule, we would see several peaks each
of which is splitted due to eventual spin-spin interaction of the
H nuclei provided by the molecular (magnecular) bonds.

The possible types of bonds and mechanisms of creation and stability of
the magnecules are:

\begin{enumerate}
\item Valence bonds (for substructures);
\item Hydrogen bonds (for substructures);
\item Polymerization (for substructures);
\item Magnetic bonds due to polarized electronic orbits;
\item Electric polarization;
\item Van der Waals bonds;
\item Three-center bonds;
\item Two-dimensional quantum tunnel effect;
\item Delocalized electrons; etc.
\end{enumerate}

Below, we consider, in a more detail, some of the possible types of
bonds in magnecule.\\

{\it 1. Valence bonds?}

In Fig.~2, we present chemical structure of a compound which
simulates presence of six CO molecules and five H$_2$ molecules.
This compound can be thought of as one of the candidates to
magnecules if we adopt the hypothesis that magnecule is due to
{\it valence} bonds between H, C, and O elements.

Indeed, here all the valence bonds of the H, C, and O have been
used thus giving support to a stability of the compound, and, this
structure can be extended by inserting additional vertical
H-O-C-C-O-H lines thus providing bigger mass of the resulting
compound, up to the 1000 a.m.u. Moreover, H-H (104.2 kcal/mol) and
C=O (255.8 kcal/mol) binding energies are bigger or comparable to
the binding energies of the valence bonds presented in Fig.~2 (see
Table~1), with the weakest bond being C-C (82.6 kcal/mol), thus
partially explaining why such a compound, when it is destroyed,
could give the detected H$_2$ and CO molecules of magnegas.

\begin{figure}[ht]
\begin{center}
\unitlength=0.5mm
\begin{picture}(100,100)
\put(10,11){\scriptsize 111} 
\put(12,31){\scriptsize 86}  
\put(12,51){\scriptsize 83}  
\put( 6,66){\scriptsize 100} 
\put(20,10){\line(0,0){7}}
\put(20,30){\line(0,0){7}}
\put(20,50){\line(0,0){7}}
\put(20,70){\line(0,0){7}}
\put(20,90){\line(0,0){7}}
\put(17, 1){H}
\put(17,21){O}
\put(17,41){C}
\put(17,61){C}
\put(17,81){O}
\put(17,99){H}
\put(40,10){\line(0,0){7}}
\put(40,30){\line(0,0){7}}
\put(40,50){\line(0,0){7}}
\put(40,70){\line(0,0){7}}
\put(40,90){\line(0,0){7}}
\put(37, 1){H}
\put(37,21){O}
\put(37,41){C}
\put(37,61){C}
\put(37,81){O}
\put(37,99){H}
\put(60,10){\line(0,0){7}}
\put(60,30){\line(0,0){7}}
\put(60,50){\line(0,0){7}}
\put(60,70){\line(0,0){7}}
\put(60,90){\line(0,0){7}}
\put(57, 1){H}
\put(57,21){O}
\put(57,41){C}
\put(57,61){C}
\put(57,81){O}
\put(57,99){H}
\put( 7,44){\line(3,0){7}}
\put(27,44){\line(3,0){7}}
\put(47,44){\line(3,0){7}}
\put(67,44){\line(3,0){7}}
\put( 7,64){\line(3,0){7}}
\put(27,64){\line(3,0){7}}
\put(47,64){\line(3,0){7}}
\put(67,64){\line(3,0){7}}
\put( 0,41){H}
\put( 0,61){H}
\put(75,41){H}
\put(75,61){H}
\end{picture}
\end{center}
\caption{Compound simulating by its content six CO and five H$_2$.
The numbers near bonds indicate corresponding average binding
energies, kcal/mol.}
\end{figure}
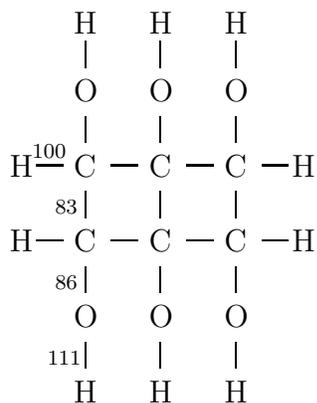

However, such a compound appears to be not relevant as a model of
real magnecules. Usually, at temperatures $T<100$C, heat
excitations of molecules are not capable to destroy the bonds with
the binding energies bigger than 30...35 kcal/mol.

At room temperature, magnegas is stable, so we can conclude that
the binding energy of the specific bonds in magnecule is estimated
to be
\be\l{BEmag}
B[magnecule] > 25...30 \mbox{ kcal}.
\ee
Due to the observed difference of the GC-MS spectra for the same
magnegas at different time of measurement \cite{1}, we conclude
that the magnecules are metastable at $T=300$K so that the
specific bonds in magnecule are partially destroyed by heat
excitations at $T=300$K. Therefore, in (\ref{BEmag}) we should
make correction to lower permitted values by taking,
conservatively,
\be\l{BEcor}
B[magnecule]>20...25 \mbox{ kcal}.
\ee

The distribution of peaks in the GC-MS spectra of magnegas tells
us that the charged fragments of magnecules are distributed
densely, with no evident order, and almost linearly increased
mass, up to some highest value of mass (e.g., 416 a.m.u.), which
can be interpreted as a mass of single-ionized magnecule. So, it
seems that that all the {\sl specific bonds in magnecule are
characterized by approximately the same value of the binding
energy}. Therefore, the magnecule does not have something like a
hard core which could be detected at a soft electron beam in
mass-spectrometer.

The infrared spectrum of magnegas \cite{1,3} does reveal presence
of the valence bonds C-H (high intensity IR), O-H (sharp IR), C-C
(weak intensity IR), and/or C-O (high intensity IR), as well as
C=O and O=C=O. However, GC-MS/IR analysis, based on about 138,000
compound data library, does not show presence of ordinary
compounds except for the associated IR peaks of carbon monoxide
and carbon dioxide. In addition, big number of strong bonds (like
valence bonds) in magnecule would require much energy to be
destroyed while, e.g., the measured anomalously high value of the
energy released at magnegas combustion leads us to supposition
that the specific bonds in magnecule are not strong. Thus, in
general we can rule out the hypothesis that the specific bonds in
magnecule are due to standard valence bonds.


It is worthwhile to note that, e.g., cobalt hydrocarbonyl,
HCo(CO)$_4$, containing both H and CO, is a gas, at room
temperatures. Such type of a compound is known as somewhat unusual
because neutral metal is bonded to carbon monoxide CO, which
mostly conserves its own properties. In a strict sense, this type
of bond is not a standard valence bond. Another example is
magnesium carbonyl, (CO)$_5$-Mn-Mn-(CO)$_5$ (melting point is
66~C), where the bond Mn-Mn is of about 40 kcal strength.

Also, it is interesting to note that nickel carbonyl, Ni(CO)$_4$,
formed at $T=80$C, is a gas at room temperatures, and dissociates,
Ni(CO)$_4 \to$ Ni + 4CO, at $T=200$C. Thus, the binding energy of
the bond between $Ni$ and CO is of about 30 kcal. We could expect
that the magnecules have a kind of hydrocarbonyl structure like
C(CO)$_n$.

It is remarkable to note that one can try to destroy magnecules by
ultraviolet light, which supplies energy of about 70 kcal/mol for
each absorbed mole of photons. However, this is the case only if
magnegas effectively absorbs the UV light.
\\

{\it 2. Hydrogen bonds?}

Since magnegas contains hydrogen the specific bonds in magnecule
could be due to the hydrogen bonds. This bond originates from H
atom mediating two other atoms by sharing its electron with one
atom and its proton shifted to the other one. The binding energy
of the hydrogen bond is about $B_{H} = 5$~kcal/mol. Examples are
HF (liquid), with the chain $\cdots \H\cdots \F \cdots \H \cdots
\F \cdots$, and H$_2$O (liquid), where the hydrogen bonds are
responsible for peculiar properties of water. We see that the
hydrogen bonds are also too weak, to meet the above estimated
value (\ref{BEmag}) of the binding energy of the specific bonds in
magnecule. However, we should to note that in the case of several
hydrogen bonds in parallel between two molecules the overall
binding energy is evidently increased.

Even if we assume that the specific bonds in magnecule are a kind
of "composite" bonds, Van der Waals bonds plus hydrogen bonds, the
estimated upper total value, $B_{VdW}+B_{H}=10$ kcal/mol, still
does not fit $B[magnecule]>20...25$ kcal. It follows from the
energy estimation (\ref{BEmag}) that the specific bond in
magnecule are weaker than typical valence bonds and stronger than
typical Van der Waals and/or hydrogen bonds, for
a structure consisting of H, C, and O atoms.\\

{\it 3. Polymerization?}

It is quite interesting to note that there is the reaction
producing an explosive compound, potassium carbonyl,
 6CO + 6K $\to$ K$_6$(CO)$_6$,
which is used to obtain unusual carbon monoxide (carbon monoxide
complex), (CO)$_6$.

This structure is believed to exist due to polymerization. For
example, a typical polymerization (e.g., of propylene,
CH(CH$_3$)CH$_2$) is owing to C-C bond, which is characterized by
the binding energy about 72...83~kcal/mol, with the reaction heat
of about $\Delta H=-20$~kcal/mol per each molecule of the linear
chain of polypropylene. We see that the typical value of the
binding energy is much bigger than 30...35~kcal/mol. However, some
mechanisms of interaction between CO molecules can also make a
contribution here because of the specific electronic structure of
CO molecule. So, we could expect lower values of the binding
energy between two neighbor CO molecules in (CO)$_6$ complex. It
is a consequence of the existence of (CO)$_6$ that there may exist
higher mass carbon monoxide complexes, (CO)$_n$, with $n>6$.

In any case, the known type of carbon monoxide complex, (CO)$_6$,
is a {\bf direct confirmation that the above conjectured
$\C\O\times \C\O$ bond really exists, as it is known in practical
chemistry.} So the existing complex
\be\l{COn}
\C\O\times \C\O\times\cdots\times \C\O\times \C\O,
\ee
where "$\times$" denote a bond, is the best known real candidate to be
a kind of magnecule. We should calculate whether the binding energy of
this bond satisfies the condition (\ref{BEmag}).

The complex of type (\ref{COn}) would give mass-spectrum which
exhibits periodicity in molecular masses of its fragments. Indeed,
the weakest bonds in (\ref{COn}) are evident so that the complex
(\ref{COn}), under the influence of the electronic beam in
mass-spectrometer, would dissociate by separating integer number
of (ionized) CO molecules. However, the mass-spectra of magnegas
does not reveal periodicity with mass 28 a.m.u. as the smallest
step. Instead, we observe almost randomly distributed masses of
the fragments, with the minimal mass difference being 1 a.m.u.

Thus, we are led to the assumption that there are some other types
of magnecules in magnegas, in addition to (\ref{COn}). We expect
presence of H atoms in magnecules which could provide (multiple)
hydrogen bonds. The -O-O- group serving, for example, as a
peroxide bridge, could be present here as well. \\

{\it 4. Magnetic bonds?}

To start with, study of the effect of external strong homogeneous
constant magnetic field, in order to see what happens with electronic
structure of (diatomic) molecule, is much important in view of the
supposed magnetic origin of the bonds.

The most interesting is to investigate the electronic structure of
many-electron diatomic molecules, especially CO (255.8~kcal/mol).
A peculiar property of the electronic structure of carbon monoxide
CO is that each of the atoms, C ($1s^22s^22p^2$) and O
($1s^22s^22p^4$), has free (lone) pair of electrons. So, only six
electrons among ten $2s$ and $2p$ electrons, namely, two electrons
from C and four electrons from O, are used to form two standard
valence $\pi$-bonds and one standard valence $\sigma$-bond, within
the framework of the molecular orbitals method. All these $\pi$-
and $\sigma$-orbitals are very close to O while the lone pair of C
is far from O. This lone pair has high energy, and can easily
react with an element that accepts electrons. So, CO reveals
electronic donor properties. This model of the CO bonds is
approximate because of the use of pure $2s$ and $2p$ orbitals;
more detailed picture can be obtained by hybridization of $2s$ and
$2p$ orbitals.

{\it 5. Van der Waals bonds?}

Three basic types of Van der Waals bonds are:

The orientational bond, which is due to the interaction of
constant dipole moments, $p_e$, of molecules,
\be\l{ori}
U_{ori} = -\frac{p_e^4}{24\pi^2\varepsilon_0^2kT}\frac{1}{r^6}.
\ee
The molecules tend to direct their dipole moments $\vec p_e$ along
one line to provide lower value of the total energy while heat
makes a disorder. The above expression for the potential energy is
valid for high temperature region.

The inductional bond, which is due to the induced electric
polarization of molecules, is characterized by the potential
energy,
\be\l{ind}
U_{ind} = -\frac{\alpha
p_e^2}{8\pi^2\varepsilon_0^2}\frac{1}{r^6},
\ee
where $\alpha$ is polarization constant. For neighbor molecules
"A" and "B", the induced dipole moment of the molecule "A" is
$\vec p_e=\varepsilon_0\alpha \vec E$, where $\vec E$ is electric
field of the molecule "B". The polarization property of the
molecule "A", characterized by a numeric value of the parameter
$\alpha$, depends mostly on the molecular volume of the molecule
"A"; for nonpolar molecules, $\alpha$ does not depend on the
temperature.

The dispersional bond, which is due to the interaction caused by
correlated zero mode oscillations of two neighbor molecules, is
characterized, in a linear approximation, by the potential energy,
\be\l{disp}
U_{disp} = -\frac{e^4\hbar\omega_0}{32\pi^2\varepsilon_0^2
a}\frac{1}{r^6},
\ee
where $\omega_0 = \sqrt{a/m}$ is frequency of zero mode harmonic
oscillations, and $a$ is coefficient of elasticity.

The above three formulas are in SI units, where
$\varepsilon_0=8.85\cdot 10^{-12}$C$\cdot$N$^{-1}$$\cdot$m$^{-2}$
is the fundamental dielectric constant, $k=1.38\cdot
10^{-23}$J$\cdot$K$^{-1}$ is Boltzman constant, $e=1.6\cdot
10^{-19}$C is charge of electron, $\hbar= 1.06\cdot
10^{-34}$J$\cdot$sec, and $r$ is a distance; 1 erg $=10^{-7}$J; 1
kcal $=4.1868\cdot10^{3}$J. Usually, the effective distance of the
Van der Waals attraction is about $10^{-9}$m $=10 \AA \simeq
20$~bohrs, between centers of molecules. For the molecules having
considerable constant dielectric moment (like H$_2$O and HCl), the
orientational VdW interaction (\ref{ori}) is dominating while for
the other molecules the dispersional VdW interaction (\ref{disp})
is a leading term. The Van der Waals bond, as a sum of the above
three types of bonds, $U_{VdW}= U_{ori}+U_{ind}+U_{disp}$, is
usually characterized by the binding energy $B_{VdW}=0.1...10$
kcal/mol.

An example is $B_{VdW}=2.4$ kcal/mol for methane ($CH_4$)
molecular crystal which has the melting temperature $T=-183$C. For
the higher mass alkanes (ethane, propane, buthane, etc.) the
melting temperature increases almost linearly, up to $T=66$C for
triaconthane (number of C atoms in triaconthane molecule is 30).
So, the Van der Waals binding energy for triaconthane molecular
crystal is estimated to be about $2.4\times(273+66)/(273-183)= 9$
kcal/mol.

>From this point of view, {\it measurement of the melting
temperature of magnegas} and {\it precise measurement of the
temperature range at which magnecules in magnegas begin to
dissociate}, at normal pressure, are of much interest, and could
give us experimental grounds for independent estimation of the
value of $B[magnecule]$, in addition to (\ref{BEcor}).

The values $B_{VdW}=0.1...10$ kcal/mol are too small, in
comparison with $B[magnecule]>20...25$ kcal/mol, to conclude that
the specific bonds in magnecule are due to the Van der Waals bonds
alone. However, in general the Van der Waals bonds can be strong
enough in some cases, e.g., for fullerene (C$_{60}$) molecular
crystal and the above mentioned triaconthane, to survive room (and
higher) temperatures. That is, $B_{VdW}$ can be of the order of 20
kcal/mol, as it is the case for fullerene crystals. So, in general
we can not rule out the possibility that the origin of the
specific bonds in magnecules are due to Van der Waals bonds.

Nevertheless, while for the molecules like C$_{60}$ or
triaconthane it is more or less justified that they provide strong
VdW bonds (about 9...20 kcal/mol) since they are molecules of big
size with big number of atoms, for H$_2$ and CO diatomic molecules
it is hardly the case; we expect the values $B_{VdW}=0.1...5$
kcal/mol for H$_2$ and CO (magnegas). So, it seems that the
specific bonds in magnecule are not Van der Waals ones alone since
they alone are too weak to fit the currently available
experimental
data. \\

Below, we consider briefly some aspects of the processes in
PlasmaArcFlow reactor, in order to give a preliminary analysis of
the conditions at which magnecules of magnegas are produced.

Strong external (electro)magnetic field may cause a deep
rearrangement of the electronic structure of many-electron
diatomic molecules, thus leading to capability to form additional
and/or rearranged bonds. On the other hand, the influence of the
external (electro)magnetic field might be a driving force in
producing of (CO)$_n$, and other gaseous molecular complexes, in
the PlasmaArcFlow reactor; notice that the constant dipole moment
of CO is about 0.12 Debay, and for H$_2$ it is zero. Indeed, the
presence of high molecular mass complexes with the specific bonds
estimated due to (\ref{BEmag}) indicates that in the reactor there
are specific conditions for {\it ordering} of elementary molecules
to the complexes despite the effect of high temperatures.
Temperature of liquid in the reactor is kept at about $T\simeq$
70C, and therefore characteristic temperatures of gas in big
bubbles are $T\simeq$ 70...100C. Very small bubbles are of high
pressure and temperature which decreases as they expand and try to
reach thermodynamical equilibrium with the liquid until they leave
it. These very small bubbles have temperature higher than
$T=100$C, and the gas in them is of high pressure. It seems that
the complexes are formed after this stage. Indeed, very high
temperatures ($T>1500$C) caused by the arc serve to dissociate
molecules while in the lower temperature region, close to the arc,
association of molecules (H$_2$, CO, etc.) takes place.

So, we indicate six characteristic temperature ranges and associated
regions:
\begin{enumerate}
\item Temperature of the underwater arc (DC, 30...40V, 500...1000A),
$T>1500$C. Dissociation of H$_2$O molecules ($\sim$~110 kcal/mol),
association of CO ($\sim$ 255 kcal/mol) and CO$_2$ molecules;

\item Temperature in the region close to arc, $800<T<1500$C.
Association of H$_2$ ($\sim$ 104 kcal/mol) and H$_2$O molecules,
the reaction (\ref{COHO});

\item  Temperature in the region near the arc, $700<T<800$C.
Very small bubbles of CO, H$_2$, CO$_2$, and H$_2$O gases, the
reaction (\ref{COHOb});

\item  Temperature in the region near the arc, $150<T<700$C.
Very small bubbles of CO, H$_2$, CO$_2$, and H$_2$O gases;

\item Temperature in the region near the arc, $100<T<150$C.
Association of O$_2$ molecules and complexes ($\sim$ 30 kcal/mol
due to (\ref{BEmag})), small and big bubbles of CO, H$_2$, CO$_2$,
O$_2$, and H$_2$O gases;

\item Temperature in the region far from the arc, $70<T<100$C.
Association of complexes ($\sim$ 30 kcal/mol due to (\ref{BEmag}),
water condensation, big bubbles of CO, H$_2$, CO$_2$, and O$_2$
gases leaving the liquid.
\end{enumerate}

Here, we have not indicated $C$ atoms which can also be present
and make an essential contribution at some of the above stages.

Microscopically, forming of the high mass metastable complexes
requires collision, or series of collisions, of several CO (and
may be H$_2$ and O$_2$) molecules, at temperatures $T<150$C.
Otherwise, the complex would be dissociated back by heat
excitations. However, we should to note that high pressure in
bubbles can admit higher temperatures, $T<200$C, or so, instead of
$T<150$C.

\section{Conclusions}

Magnegases are essentially composed of metastable clusters, called Santilli's
magnecules [1c], containing isolated H, C, and O atoms plus OH, CH dimers,
plus H$_2$ molecules, single, double, and triple CO bonds.

The initiation of the combustion breaks down those clusters after
which some of H atoms form H$_2$ molecules while other form H$_2$O
without forming H$_2$ molecules. It is important to note that
prior to combustion, only a small number of C and O atoms are
combined into carbon monoxide. As
established by experimental evidence, CO$_2$ is contained in the
exhaust in about 5\%. This evidence is particularly intriguing in
view of the fact that 50\% of magnegas produced from pure water
and carbon electrodes must be composed of C and O atoms. The fact
that the combustion exhaust of the same gas contains about 5\%
CO$_2$ therefore establishes beyond doubt that only a small
percentage of C and O atoms are bonded into carbon monoxide. Still
in turn this evidence establishes magnecular structure of
magnegas.

In view of all the above results the combustion of magnegas can be
optimized by the following means:

1) Magnegas combustion should run at lower temperatures as
compared to other fuels. In particular, this implies automatical
reduction of NOx's;

2) Combustion of magnegas should be triggered by spark operating
at the highest possible voltage with a minimum of the order of
50,000~V;

3) Energy content of magnegases can be increased with
increase of operating pressure and electric power.

In closing, we should indicate that the number of open problems on
the creation, structure and combustion of magnegas has increased,
rather than decreased, following this study. This outcome should
have been expected in view of the novelty, extreme complexities,
as well as insufficient experimental data at this writting
on the magnegas technology.

It is hoped that this study will stimulate additional research by
interested colleagues in view of the important environmental
implications of this new fuel.

\end{document}